\documentclass[9pt,letterpaper,onecolumn]{sig-alternate}
\usepackage[T1]{fontenc}
\usepackage[latin9]{inputenc}
\usepackage{xcolor}
\usepackage{graphicx}
\PassOptionsToPackage{normalem}{ulem}
\usepackage{ulem}

\makeatletter

\providecolor{lyxadded}{rgb}{0,0,1}
\providecolor{lyxdeleted}{rgb}{1,0,0}

\usepackage{balance}
\numberofauthors{3}
\author{
\alignauthor 
Rute Sofia, \\ Paulo Mendes\\
\affaddr{SITI,University Lusófona, Portugal\\
rute.sofia@ulusofona.pt \\ paulo.mendes@ulusofona.pt}
\alignauthor 
Manuel José Damásio, \\ Sara Henriques\\
\affaddr{CICANT,University Lusófona, Portugal\\
 mjdamasio@ulusofona.pt\\ shenriques@ulusofona.pt}
\alignauthor 
Fabio Giglietto, \\ Erica Giambitto\\
\affaddr{Department of Communication Studies, University of Urbino\\
fabio.giglietto@campus.uniurb.it\\ e.giambitto@campus.uniurb.it}
\and
\alignauthor 
Alessandro Bogliolo\\
\affaddr{STI-DiSBeF, University of Urbino\\ alessandro.bogliolo@uniurb.it}
} 
\additionalauthors{}
\balance

\let\@copyrightspace\relax

\makeatother

\begin{document}

\title{Moving Towards a Socially-Driven Internet Architectural Design}
\maketitle
\begin{abstract}
This paper provides an interdisciplinary perspective concerning the
role of prosumers on future Internet design based on the current trend
of Internet user empowerment. The paper debates the prosumer role,
and addresses models to develop a symmetric Internet architecture
and supply-chain based on the integration of social capital aspects.
It has as goal to ignite the discussion concerning a socially-driven
Internet architectural design. 
\end{abstract}

\subsection*{Categories and Subject Descriptors}

A.1 \textbf{{[}}Introductory and Survey\textbf{{]}; }C.2 \textbf{{[}}Computer
Communication Networks\textbf{{]}: }Network architecture and design. 
\begin{keywords}
Internet architectural design; Future Internet foundations; prosumer;
social capital; knowledge; supply-chain.
\end{keywords}

\section{Introduction}

\label{sec:intro}

The Internet has reached a new era in its evolutionary track, an era
where user empowerment and engagement are gaining momentum due to
the wide availability of the most varied \emph{Social Media}. Social
Media are often solely associated to \emph{Online Social Networks
(OSNs)}, to \emph{Social Network Sites (SNSs)\cite{boyd2007social},
}or to Web-based tools, such as Flickr, that allow Internet users
to exchange content. Within the context of mobile networks, Social
Media is the umbrella for mobile software that allows Internet users
to actively engage in sharing some of their interests in their daily
life experience. In the context of new media, Social Media expands
to embrace any form of digital media where users can actively engage,
e.g. digital Television, while if we consider Internet architectures
and their evolution, then Social Media entails also disruptive architectures
which rely on some form of user cooperation as occurs with \emph{User-centric
Networks (UCNs)} \cite{consortium2010eu}, i.e., networking architectures
which grow in a ``viral'' way through user engagement, exchange
of shared interests, and cooperation incentives. \textbf{In this paper,
we refer to Social Media as the whole set of tools aforementioned.}

The fast-paced adoption of Social Media seems to provide Internet
stakeholders with new forms of expression and knowledge exchange.
Yet, from a provider's perspective these tools are still exclusively
seen as a part of digital marketing. However, new results derived
from the use of Social Media - such as open-data sets - are indicators
towards the need to revisit Internet architectural design and to provoke
a shift towards a more symmetric Internet supply-chain model by incorporating
both economical and societal (interaction) aspects.

This paper is dedicated to the debate about the need to take an interdisciplinary
approach to the design of future Internet architectures, services,
and technologies. The paper is organized as follows. Section \ref{sec:RelatedWork}
goes over related work, while section \ref{sec:Prosumer} addresses
the prosumer notion from the perspective of different layers on the
OSI stack. Section \ref{sec:Internetvaluechain} gives a perspective
on the impact of the prosumer role integration into Internet supply-chain
models. In section \ref{sec:architecturaldesign} we provide guidelines
on the evolutionary process to drive the Internet architectural design
into a structure that truly embraces both a technological and societal
perspective more suitable to incorporate a \emph{prosumer} notion,
being the paper concluded in section \ref{sec:Conclusions}.

\section{Related Work}

\label{sec:RelatedWork}

The need to revisit Internet architectural design has been almost
a constant on the past decade, from a technological perspective. Projects
such as GENI \cite{foundationglobal} as well as initiatives such
as EIFFEL \cite{EIFFEL} have given rise to a wide variety of innovative
technological aspects concerning Future Internet foundations.

In the most recent years there has also been an increase in multidisciplinary
work within the context of social networking and social interaction
analysis. SocialNets \cite{consortium2008socialnets} addresses the
evolution of social structures relying on a cross-field perspective
which combines pervasive networking and some aspects of social sciences
related to human behavior. SocialNets lead to a better understanding
of metrics (derived from a human interaction behavior) as well as
to a better comprehension of assumptions that are being used to model
our perception and knowledge of Internet evolution, e.g., the way
that mobile nodes move and the way that users behave. 

S. Ferlander analyzed the potential impact of technology on the development
of social capital and of new communities in urban environments \cite{ferlander2003psychology}
both in terms of knowledge generation and in terms of community attractiveness.
Pénard and Poussing have analyzed and formulated several hypothesis
concerning Internet use and the development of social capital \cite{penard2008internet},
having revealed that there is a significant positive impact both in
terms of increase in volunteer activities and in terms of trustworthiness
in online investments, concerning well established social capital
ties. 

A third relevant field of work to be cited is the one of social networking
analysis as a multidisciplinary effort that is being applied across
several domains. Borgatti et al. provide a multidisciplinary perspective
to such evolution by describing how social networking structures \cite{everett2005ego}
and related definitions can be addressed from a supply-chain modeling
perspective \cite{borgatti2009onsocial}.

\section{User Empowerment Models}

\label{sec:Prosumer}

This section addresses two main user empowerment models that we believe
impact Internet foundations design: the \emph{prosumer} model, which
is based on the need for collective expression; and the \emph{network
prosumer} model, a natural evolutionary step of user empowerment and
participatory models which, allied to the pervasiveness and lower
costs of networking technology, is giving rise to new types of Internet
architectures.

\subsection{Prosumer}

It has often been debated that the current Internet end-user is moving
from a plain consumer towards a \emph{prosumer }model \cite{toffler1980bantam}.
The key difference between a consumer and a prosumer is that the latter
embodies a form of empowerment in the sense that the user plays an
active role in improving products/services. 

The impact of this model on Internet architectural design relates
to the real value of Social Media. Such value is created by users
who share their interests (their perspective of knowledge based on
their daily life experience) in the form of digital content - user
generated content. This sharing seems to go beyond a basic human need
of socialization, being driven by a social need related to \textbf{collective
expression}. 

Ritzer et al. have further analyzed the prosumer notion, having considered
a fast-food metaphor to describe a change that Ritzer first observed
in the American society \cite{ritzer2010mcdonaldization}. Ritzer
then pursued an analogy concerning the role of Social Media, in the
form of OSNs and SNSs \cite{ritzer2010mcdonaldization}, pointing
out negative aspects such as the possibility of abuses in the form
of unpaid labor. This unpaid labor is, according to the authors, only
in part balanced by the fact that Social Media are often offered for
free.

The prosumer model is also impacting content dissemination strategies.
Social Media give the means to disseminate knowledge in new ways and
based on new formats; the Internet user has the means to enjoy services
in a quite independent way. Even more relevant is the fact that Social
Media allow the consumer to became a producer at a quite low cost.
Hence, \emph{convergence} becomes a product of social, cultural, industrial,
and technological changes; a process that influences and modifies
the circulation of knowledge (culture) \cite{jenkins2006new} to create
a collective expression (social) by means of even more complex and
pervasive products (technology, industry) \cite{ito2008innetworked}.

A final relevant aspect related to the prosumer model is that its
Internet presence seems to exhibit power-law properties \cite{2003barab225si,huberman2001laws}.
Such properties are a sign of the presence of a social process in
OSNs, as a small percentage of users produces the most significant
share of content, while the majority remain within the consumer model;
these observations provide the proves of the coexistence of the consumer
and prosumer models and their impact on the evolution of the Internet
design.

\subsection{The Network Prosumer}

Today, the prosumer notion embraces more than content development
and user engagement/participation: it impacts Internet access provisioning,
as the Internet end-user has at his/her disposal technology that allows
him/her to behave as a \emph{network prosumer. }In \emph{UCNs} \cite{consortium2010eu,sofia2008userprovided},
the user becomes actively engaged on the networking operation and
process. From an Internet connectivity model perspective, a UCN can
be represented as a time and space varying graph where nodes are wireless
devices belonging to Internet users, and where edges represent trust
and affinity associations. The edge cost is a measure of the trust
association strength as well as the level of influence that users
play on each other. From a pure connectivity perspective, nodes have
two roles: \emph{regular} and \emph{network prosumer} (NP). Regular
nodes use network resources provided either by an NP or by a regular
access provider. The NP is therefore a prosumer at a networking level.
NPs provide (networking) services to a specific community of users,
e.g. share bandwidth driven by a social process (shared interests,
even if amongst real-life strangers) or in a coordinated way with
one or several access providers.

UCNs reside on the Internet fringes and this is a consequence of prosumer
integration. Prosumers rely on low-cost wireless technologies, software
defined networks, and also on their willingness to cooperate due to
some form of communal or individual benefit (\emph{incentive}) - \textbf{a
social aspect}. Moreover, UCNs embody four main properties: \emph{network
resource sharing}, \emph{cooperation}, \emph{trust}, and \emph{self-organization}.\emph{ }

\emph{Network resource sharing} today is mostly associated to Internet
access or connectivity sharing. However, as these architectures evolve,
we will likely observe sharing of additional network resources (e.g.
energy) or of additional network services (e.g. mobility management). 

\emph{Cooperation} relates to users willingness to participate in
UCNs, both sharing and profiting from available resources. Incentives
to cooperate can be related to trust (e.g. social association), to
some form of gratification (e.g. broader Internet access), or even
to a more efficient network operation. 

\emph{Trust management} is today performed by having users signing
up to a ``community''. However, to create UCN secure environments,
user identification and traceability are issues that have to be addressed.
Hence trust management relates to three main concerns:\emph{ i) assist
users in terms of traceability; ii) guarantee user privacy; iii) provide
data confidentiality when/if necessary}.

\emph{Self-organization} relates to the capability to coordinate connectivity
in scenarios where\emph{ }it is based on users willingness to cooperate
or adhere. 

The example based on UCNs intends to explain why, from our perspective,
the prosumer role is moving towards the lower layers of the OSI stack
thus eventually resulting in significant implications on the Internet
architectural design.

\section{Prosumer Impact on the Internet Supply-Chain}

\label{sec:Internetvaluechain}

This section provides a perspective on the Internet supply-chain evolution
and also debates on how we believe the Internet supply-chain may evolve,
based on user empowerment and from a technological perspective, towards
a stage that is more prone to consider knowledge exchange as \emph{Return-of-Investment
(RoI)}.

\subsection{Asymmetry in the Internet Supply-Chain}

\begin{figure}
\includegraphics[scale=0.28]{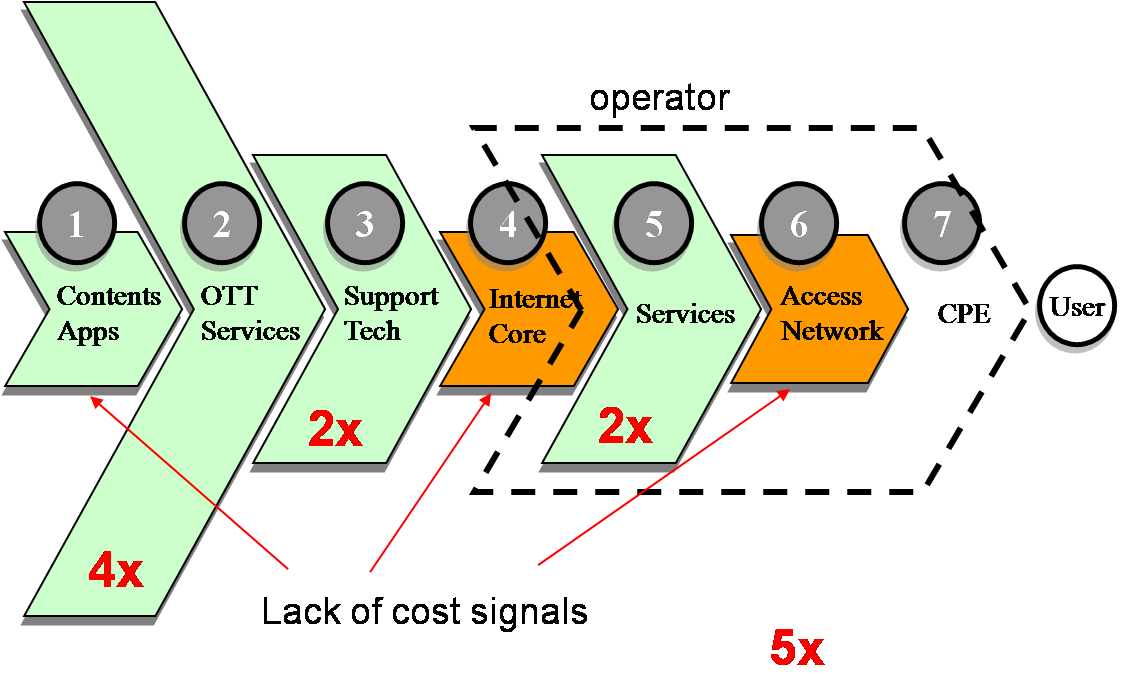}

\textbf{\caption{\label{fig:Internet-supply-chain.}Internet supply-chain representation.}
}
\end{figure}

User empowerment requires a paradigm shift in Internet architectural
design as a way to unlock the potential of new business models, which
can only be deployed if the Internet supply-chain becomes symmetric.
To explain this perspective let us consider Fig. \ref{fig:Internet-supply-chain.},
where a 7-stage model for the Internet supply-chain \cite{pigliapoco2011a}
is illustrated. In this model the Internet is seen as a two-sided
market where bits are the product unit, and where users are seen as
consumers. The supply chain follows a producer to consumer flow comprising
seven stages: (1) content and application right owners; (2) \emph{Over-the-Top
(OTT)} online services; (3) support technology, e.g. hosting services
and content delivery networks; (4) Internet core, made of exchange
points and core networks of incumbent operators; (5) managed services
directly provided within operators\textquoteright{} networks; (6)
access networks; (7) customer premises equipment and software components
used to connect to network termination points, and to gain access
to the Internet. The dashed line in Fig. \ref{fig:Internet-supply-chain.}
shows that operators tend to adopt a vertically integrated business
model in order to use the profits generated by thriving market segments
to sustain the stagnating ones. Vertical integration, however, does
not provide an ultimate answer, since it contrasts with the modular
nature of the TCP/IP stack; it limits innovation; and it creates a
misalignment between costs and price models. 

It is apparent that some of the stages (2, 3, 5, and 7) have taken
advantage of the exponential growth of Internet traffic, while some
others (1, 4, and 6) have suffered from the lack of price signals,
which has prevented investors from supporting their development. For
instance, the \emph{Capital Expenditures (CapEx)} required to increase
the capacity of fixed and mobile networks at the rate of IP traffic
growth are much higher than those estimated by projections of historical
data. The bottlenecks created by stagnating segments risk the impairment
of the development of the Internet as a whole, unless new models are
adopted. 

\textbf{}

\subsection{Symmetric Internet Supply-Chain, Accommodating  Prosumers}

\begin{figure}
\includegraphics[scale=0.35]{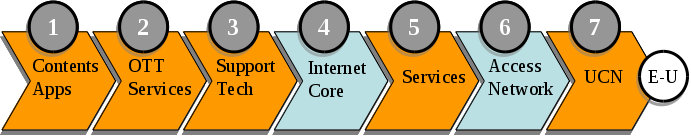}

\caption{Supply-chain with the Internet user (E-U) as recipient.\label{fig:Supply-chain-impact.}}

\end{figure}

Accommodating prosumers implies the need for symmetry, from a supply-chain
perspective. Fig. \ref{fig:Supply-chain-impact.} provides an illustration
of the Internet supply-chain previously depicted in Fig. \ref{fig:Internet-supply-chain.},
having as recipient the Internet user (E-U). For the sake of clarity,
we provide concrete technological examples for each of the stages.
For instance, (1) could be a community where some Internet service
is shared, OSNs, or applicational markets (e.g. Android market); (2)
could refer to blogs, Wiki pages, or online co-working initiatives;
(3) could integrate new forms of collective networking such as Peer-to-Peer
or overlay networks; (5) could embody community-scale networks and
services; (7) could represent UCNs extending and complementing access
infrastructures.

Fig. \ref{fig:Supply-chain-impact.} introduces three new features
when compared to Fig. \ref{fig:Internet-supply-chain.}: firstly,
it points out the impact of users on all of the different stages;
secondly, it represents UCNs as part of the supply-chain; thirdly,
it integrates the prosumer as part of the supply-chain. What Fig.
\ref{fig:Supply-chain-impact.} intends to exemplify is that when
we speak of the need to remove asymmetry, we are referring to services
and to their time correlation; not to deep, clear-slate foundational
changes as well shall explain in the next section.

\section{Socially-Driven Internet Design}

\label{sec:architecturaldesign}

Albeit social networking is currently a research area that is being
addressed by the most varied fields (e.g. pervasive networking; social
sciences; communication sciences; behavioral economics), there is
also a concrete need to align the different notions and perspectives
derived from research on these fields. For instance, most of the solutions
in the field of opportunistic routing relies today in several social
similarity notions, being \emph{centrality} one of the most used social
similarity concepts applied. However, there are strong discrepancies
concerning centrality depending on its applicability domain. Specifically,
new Internet services/solutions are taking into consideration notions
from complex networks, while leaving aside societal aspects which
can be derived from the application of social capital models and metrics,
i.e., an integrated perspective of the different, multidisciplinary
fields. For instance, to derive networking operation based on a behavior
that is closer to the one of social structures, social-based opportunistic
routing solutions should consider the dynamism of users\textquoteright{}
behavior and affinity resulting from their daily routines in order
to create durable networks in dynamic scenarios \cite{Opo12}. In
the future, Internet graph representations should incorporate information
about the affinity level of user profiles (e.g. based on behavior
and knowledge). To achieve this, it is necessary to adopt metrics
that can track social capital evolution, and not only network centrality
metrics (degree, betweenness, closeness), as the latter only assist
the coordination of actions within a social structure.

This contribution alerts to the need to make the Internet architectural
design evolve by adequately integrating user empowerment and that
such compliance goes well beyond the need to address new business
models. Instead, it is necessary to address social structures evolution
and to embed new lines of thought, multidisciplinary in nature, which
can then give rise to the development of metrics and algorithms that
may truly sustain not only a self-organizing and potentially power-law
based nature of the Internet, but also its evolution towards a robust
knowledge exchange platform. This can be achieved if technological
adoption techniques coexist with societal adoption metrics. For the
latter, social capital models and metrics are the relevant embodiment
in terms of application to the Internet architectural design.

\subsection{Brief Introduction to Social Capital}

Social capital is a concept that today is applied in a wide variety
of fields, e.g. economy, media studies, sociology. Its roots can be
traced back to Bourdieu, who defined social capital as \textquotedbl{}the
aggregate of the actual or potential resources which are linked to
possession of a durable network of more or less institutionalized
relationships of mutual acquaintance and recognition\textquotedbl{}
\cite{bourdieu1986forms}. To give a concrete example related to social
capital applicability on the field on Future Internet architectural
design, ``potential resource'' can be seen as the knowledge exchanged/gained
through interaction with specific clusters of nodes (e.g. affiliation,
family); ways to measure such resources can be ``gratitude'' or
``trust''. 

As Bordieu's definitions have been hard to quantify across different
fields, Coleman's definition of social capital \cite{coleman1988social}
gained wider acceptance. Coleman defines social capital as a set of
``entities'' that share two main features: i) each entity is part
of a social structure ii) each entity has, on that social structure,
a concrete purpose which facilitates some interaction among individuals
belonging to the structure. Coleman has also contributed with potential
models to derive social capital, but has been often criticized for
failing in contextualizing relationships and structures in a larger
socioeconomic perspective. Putnam's notion of social capital is the
most popularized one \cite{putnam2001bowling}. He defines social
capital as a set of features that assist in facilitating and coordinating
actions in structures. From a social networking perspective, examples
of such features can be levels of trust, or reciprocity. Recent conceptions
of social capital perceive it as a metaphor about the advantage that
is inherent to the strength of social relationships, and the access
an actor has to the resources available in a network \cite{bourdieu1986forms}.
This concept is an aspect or function of a social structure, or it
refers to resources embedded in a social structure \cite{burt2000network}. 

Although from a networking perspective social capital is still a notion
that is far from being quantified, there seems to be a common link
to social networking: associations between nodes (relations) make
a difference from the global network perspective. In other words,
there is a concrete relational notion associated to social capital,
that we believe is essential to consider when developing novel networking
structures and architectures. Moreover, different dimensions of social
capital can be delineated in function of several elements. 

Furthermore, social capital value can be addressed both from an \emph{individual}
as well as from a \emph{collective} perspective. On the individual
level \cite{bourdieu1986forms}, individuals are entitled (based on
reciprocity) to claim access to resources possessed by other members
of the network (cluster, community). The amount of social capital
to which an actor has access to depends on both the quantity of the
network connections that he/she can enlist, and the sum of the amount
of capital that each network member possesses \cite{glanville2009typology}.
Individual level social capital claims stress the ability of the actor
to secure benefits via social structures, suggesting that social capital
is to be regarded as social resources that are accessible through
participation in various types of social networks \cite{rostila2011facets}.
However, the process of making individual resources available to others
through social relationships does not assume that social capital is
solely \textquotedbl{}owned\textquotedbl{} by individuals. As Coleman
suggests, social capital, unlike other forms of capital, is inherent
to the structure of relations between individuals. This difference
between actual and potential resources \textendash{} the ones that
individuals use and the ones that are available on the network \textendash{}
assumes the previous existence of a relation as a condition for social
capital to be used. That is why the term \textquotedbl{}individual
social capital\textquotedbl{} is in fact misleading, as social capital
is always relational, although it can be used to achieve individual
ends. 

The collective social capital definition is provided by Putnam who
claims that social capital is created through citizens' active participation
in organizations and groups, but is in itself a set of features of
social organizations \textendash{} like trust, norms, and structures
\textendash{} that can help, via coordinated actions, in creating
a better society \cite{putnam2001bowling}. Trust is central in Putnam's
notion of social capital. He emphasizes both \emph{formal} (i.e. participation
in organizations) and \emph{informal} (i.e. socializing with friends)
collective expressiveness, where social capital is a collective good,
one that is non-exclusive in terms of consumption and that is publicly
available, though a part of the knowledge remains unleashed and intrinsic
to the structure of social relationships.

\subsection{Social Capital Metrics}

Within the context of social capital, the Internet can be contextualized
as a complex structure composed of network clusters. Each of these
clusters contains a set of nodes (\emph{actors}) that are linked together
by edges (\emph{relations, associations}), whose cost can be derived
both from virtual and from real-life interaction. From a networking
perspective we highlight that interactions may occur between nodes
that do not have a relation in real-life, e.g. strangers traveling
on the same bus. The network evolves as the actors develop some kind
of link (a single type of relation), either formally or informally.
Metrics related to these associations can be e.g. friendship; trust;
influence; recognition; reciprocity; knowledge. These are therefore
metrics derived from social behavior, and knowledge gained/exchanged
seems to be the common link for both formal or informal associations,
where actors exhibit some form of ``shared interest''. Being capable
of quantifying these metrics up to some extent is a key aspect to
develop algorithms that can assist the Internet to evolve into a robust
knowledge-generation architecture, and social capital seems to have
a primordial role in such evolution, as it integrates principles that
can be used to facilitate knowledge exchange. For instance, such principles
can assist in providing sustainability required in the infrastructure
to foster the interactions and beliefs that feed the commendable cycle
of connectedness and trust/reciprocity, both integrating positive
and negative outcome. A concrete example of negative outcome that
can be embodied relates to alienation due to the heavy usage of Social
Media \cite{putnam2001bowling}. It is also relevant to consider models
that address the Internet as a means for fostering interaction \cite{blanchard2004effects,katz2002social,ling2008new},
engagement, and social activism \cite{haythornthwaite2005social}.

\begin{figure}
\includegraphics[scale=0.3]{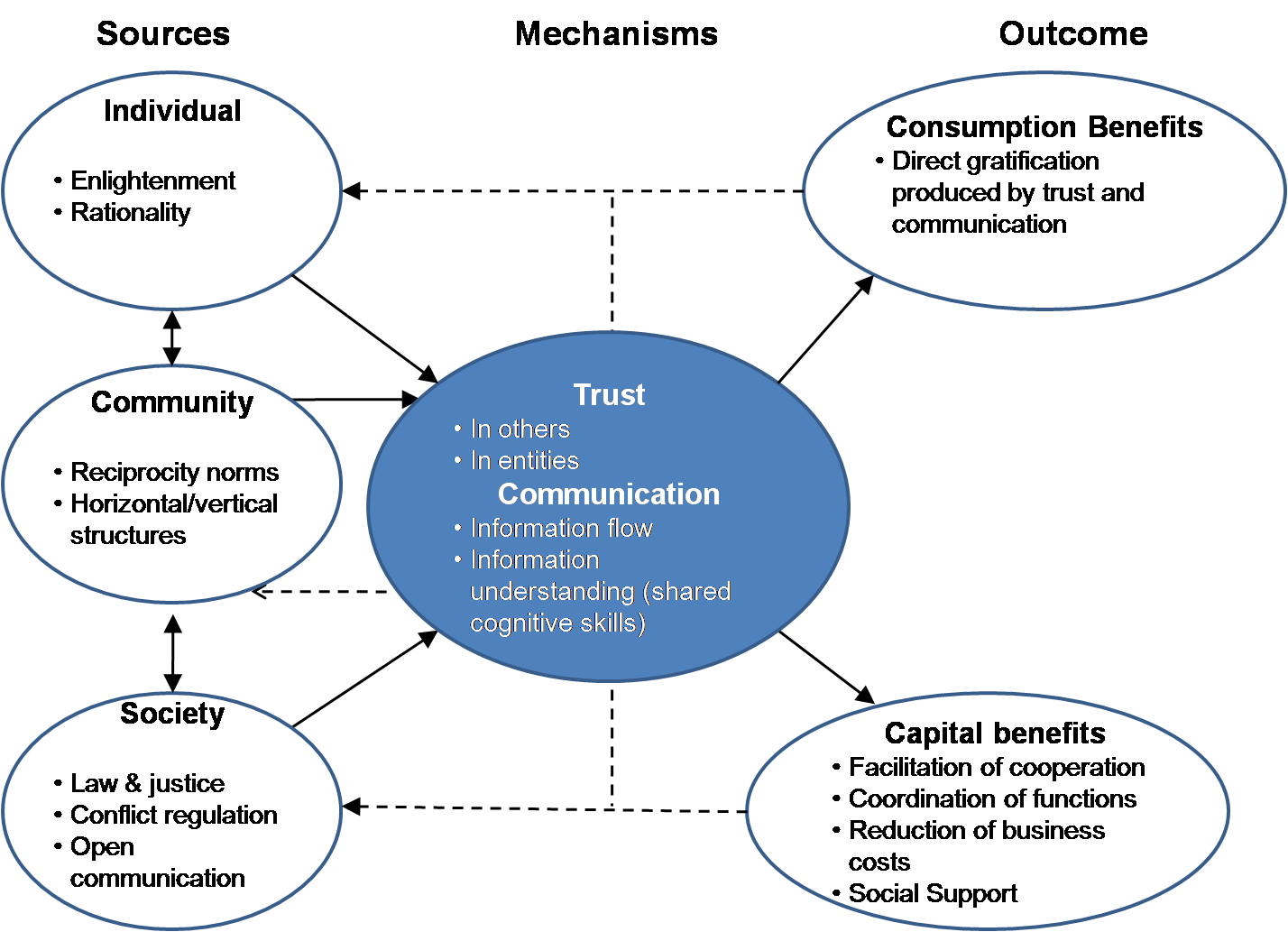}

\caption{\label{fig:Sources-and-outcome}Sources and outcome of social capital.}

\end{figure}

Fig. \ref{fig:Sources-and-outcome} provides a global perspective
concerning the mechanisms of social capital creation and derived outcome,
having as means both communication and trust. Trust stands for participation
while communication stands for the interactions derived from Social
Media usage. If we take a more network oriented approach, we would
have - as a basic source of social capital - interactions at an individual
level. If we take a more social approach, we would consider trust
as the core element prompting communities and society in its entirety
to act and develop ties. On the other side of the diagram we have
consumption benefits as a direct outcome of individual interactions,
and capital benefits as a more collective outcome.

Based on Fig. \ref{fig:Sources-and-outcome} we can verify that the
sources of social capital integrate both an individual and collective
relational approaches; the central mechanisms are structural and cognitive
forms of relationships, and the outcome is either of an individual
or of a collective nature. This perspective emphasizes trust and reciprocity
as core elements and look at participation as an outcome of the process.
If, on the contrary, we were to focus on resource availability, we
would then have the media as an internal element of the system intrinsic
to the relation between the individuals and the gratifications derived
from consumption. These two distinctive views assign to technology
completely different roles as either enablers or blockers of social
capital. Therefore, social capital theory presents a set of definitions
and relationships that seem valuable to be integrated into technology
adoption modeling of the Internet, to be able to drive the Internet
towards (also) a socially-driven design and as consequence, towards
a more robust platform for knowledge generation and exchange.

\subsection{Why Do we Need Social Behavior?}

The pervasive adoption of Social Media is a strong indicator of the
need to revisit the Internet supply-chain to be able to truly take
advantage of the prosumer model. Being capable of both integrating
social metrics and technological adoption metrics into multi-objective
utility functions is a requirement to further evolve the Internet
supply-chain. Such a methodology requires as a first step to go beyond
the pure technological perspective and to incorporate the social capital
perspective. Today, it is widely accepted that the Internet end-to-end
design principle \cite{clark1988design} is hedged around with stronger
caveats than before. Hence, we must be open to understand how we can
establish design processes that allow evolution towards the future
requirements without adding further entropy to the natural Internet
evolution process.

\begin{figure}
\includegraphics[scale=0.3]{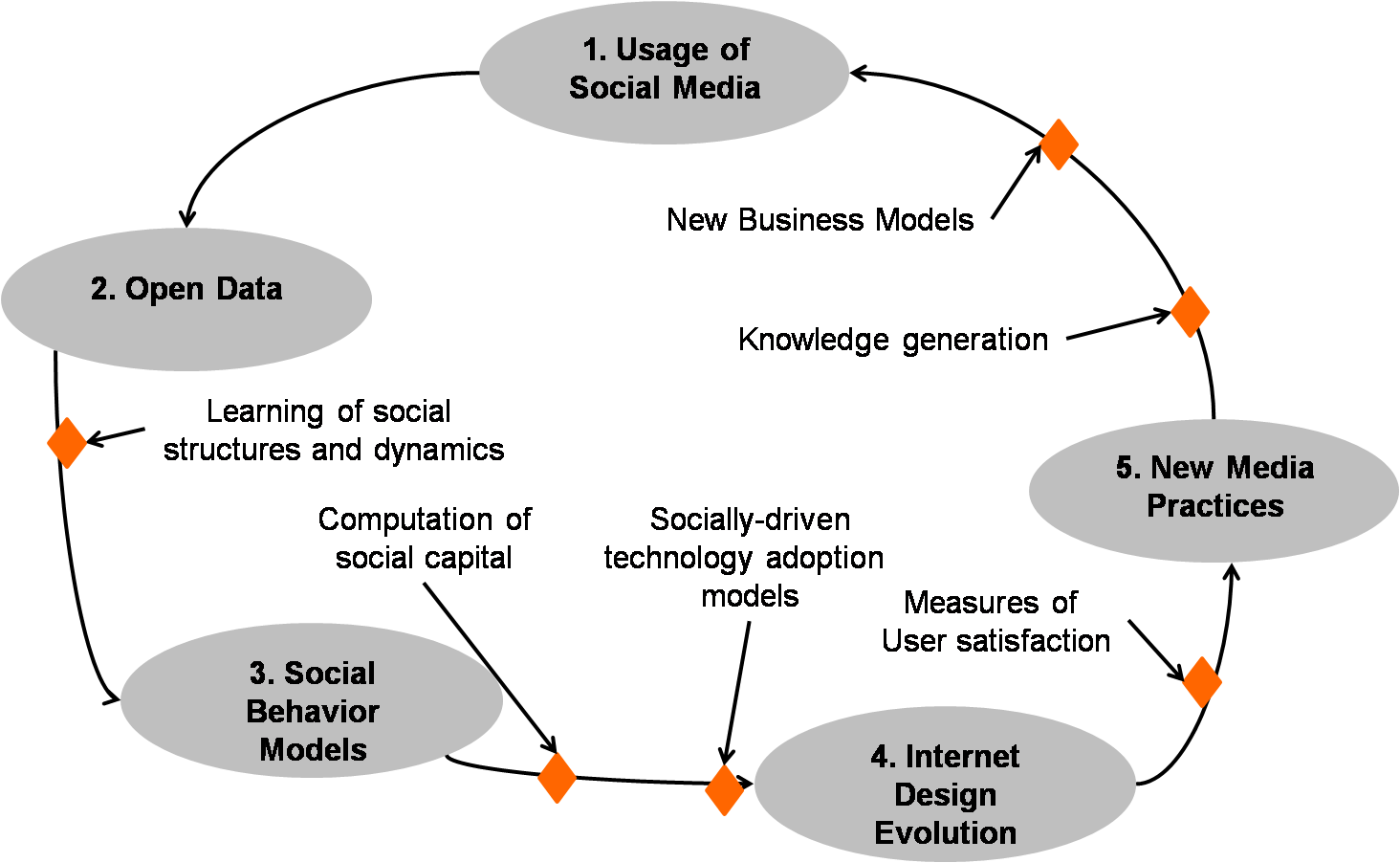}

\caption{\label{fig:socialdesign}Socially-driven design cycle.}
\end{figure}

Fig. \ref{fig:socialdesign} illustrates our proposal for the evolutionary
and gradual development of a Future Internet. Social Media today are
already the main tools assisting the development of large-scale open-data
sets (1., 2.). These in turn are more and more the study basis for
social structures dynamics. An Internet architecture aware of social
behavior models (3.) will give rise to new media practices, as well
as new business models and knowledge creation (4.). Devised social
behavior models are useful for contextualizing relationships and structures
in a larger socioeconomic perspective, aspect which is relevant for
the computation of social capital as well as for identifying features
that assist in facilitating and coordinating actions in structures
(5.). 

In order to allow the Internet design to be adjusted based on evolving
social models,  behavior of  a node should  express the logic of its
computation without describing its control flow. Hence, the design
of future Internet functionality should be based on an expressive
language (e.g. declarative or functional programming) in order to
accommodate a potentially more complex event structure and node operation.

\subsection{Incorporating Social Capital into Internet Design}

Within the context of social capital, it is our belief that the social
properties that are the most relevant to be applied to Internet design
are \emph{reach, engagement,} and \emph{influence}.

\begin{table*}
\caption{\label{tab:Centrality-definitions}Centrality definitions, the social
capital and the networking perspectives.}

\center

\includegraphics[scale=0.4]{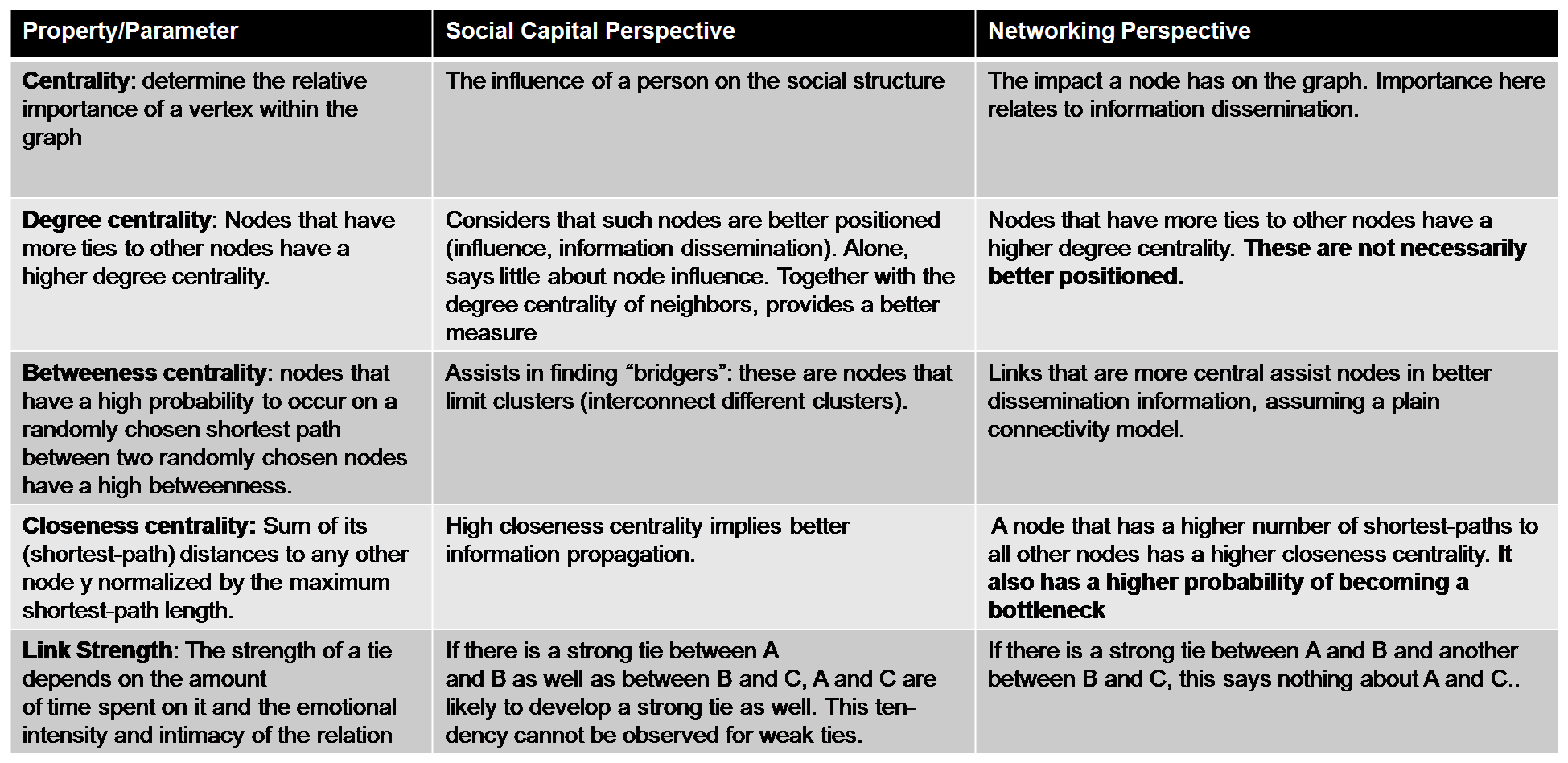}
\end{table*}

\emph{Reach} corresponds to the degree of effective dissemination
of certain content or potential spread that a single actor or node
(a profile) has in the network. Networking measurement metrics that
can be applied to incorporate this property may be, for instance,
rate of nodes reached; proximity; propagation speed.

\emph{Engagement} refers to the degree of participation and involvement
of a specific actor or node. A profile in networking can be seen as
e.g. a preferred location; an interest towards a node/cluster/location.
Metrics that we can consider in Internet architectural design to define
engagement can be, for instance, the growth of the direct neighbors
(also known as \emph{followers}) of nodes; time spent around a specific
node (e.g. volume of inter-contact times). Reciprocity of contacts
is also a metric that can assist in defining engagement.

\emph{Influence} refers to the degree of attention and mobilization
that a certain actor can generate in other actors. From a networking
perspective, influence is by far the hardest property to quantify.
A potential approach to such quantification is provided in FRINGE
\cite{FRINGE} for the context of community detection. Such proposal
can be a starting point to attempt to quantify the notion of influence
in pervasive networking. We highlight that measuring up to some extent
influence is not to be mistaken by metrics that measure node popularity.
The integration of influence is expected to assist in a better definition
of interaction matrices, an aspect that today is key for several aspects
of the operation of the Internet, such as new ways to route traffic,
or a better definition of self-organizing environments.

In addition to incorporating new metrics rooted on social science,
it is also necessary to revise a few aspects concerning network centrality.
Today, several notions of centrality are the basis for new concepts
being addressed in the Internet, e.g. information-centric routing;
opportunistic routing; self-organization based on small-world evolution.
However, there are a few differences between the application of centrality
as it is being done today in the context of networking, and within
the social capital modeling context. In Table \ref{tab:Centrality-definitions}
we provide the two perspectives for the most popular centrality definitions
being applied in Internet architectural design. 

As shown in the table the definition for \emph{degree centrality}
when applied to the context of social capital modeling differs from
the definition being employed in networking: in networking, nodes
with a higher degree centrality are not necessarily better positioned
in a network. 

Crucial differences arise also in the application of \emph{betweeness
centrality}. When applied to social capital modeling, nodes with a
higher betweenness centrality are cluster heads known as \emph{bridgers:
}their power resides in assisting in the interaction between different
clusters. However, within the context of networking nodes with a higher
betweeness centrality supposedly assist in a better dissemination
of information as they are more ``central''. The role of bridger
is not addressed from a networking perspective and yet, this is a
highly crucial role as it assists the dissemination of information
across different communities.

We observe also some discrepancies in the notion of\emph{ closeness
centrality}. While in social capital models nodes with a higher closeness
centrality imply better dissemination of information, in networking
such nodes will most likely end up being bottlenecks.

A final property that relates to the notion of \emph{link strength}
(e.g. trust association) incorporates within the context of social
capital modeling the transitivity property: if A and B share a strong
link, and B and C share also a strong link, then it is highly likely
that A and C shall also share a strong link. This is currently not
incorporated in pervasive networking, from an information dissemination
perspective.

Summarizing, it is our belief that a starting point to address a socially-driven
Internet design can simply start by addressing two simple aspects:
i) integrate the notions of trust and influence in pervasive routing,
by developing measurement metrics capable of sustaining such properties;
ii) revise the notions of centrality that are being heavily applied
today, ensuring that there is alignment between the definitions that
are today applied in exclusivity within the context of social capital,
and in pervasive networking.

\section{CONCLUSIONS}

\label{sec:Conclusions}

This paper addresses the need to consider a real merging of social
capital principles into technology adoption modeling as a way to assist
future Internet architectures to naturally evolve beyond their role
for service provisioning, thus enabling network prosumer models to
be fully exploited as tools that can give rise to new business models
and to both social and technological advances. 

Our belief is that this is a process that can be applied to the natural
evolution of the network core, by removing artificial barriers related
to Internet supply-chain management, as well as by incorporating a
multidisciplinary perspective to the dynamics of social structures,
through the integration of social capital models and metrics. To assist
in such integration, we have provided a few design guidelines concerning
how the implementation of such changes could be applied to the current
Internet architecture.

\bibliographystyle{acm}
\bibliography{fia2012}

\begin{thebibliography}{10}

\bibitem{2003barab225si}
{\sc {Barab\'{a}si, A. L.}}
\newblock {\em Linked: {How} everything is connected to everything else and
  what it means for business, science, and everyday life}.
\newblock New York: Plume, 2003.

\bibitem{blanchard2004effects}
{\sc Blanchard, A.}
\newblock The effects of dispersed virtual communities on face-to face social
  capital.
\newblock In {\em Social Capital and Information Technology}, H.~M. and W.~V.,
  Eds. Massachussets, MA, 2004.

\bibitem{borgatti2009onsocial}
{\sc Borgatti, S., and Li, X.}
\newblock On social network analysis in a supply chain context.
\newblock {\em Journal of Supply Chain Management. Volume: 45, number 2\/}
  (2009).

\bibitem{bourdieu1986forms}
{\sc Bourdieu, P.}
\newblock The forms of capital.
\newblock In {\em Handbook of theory and research for the sociology of
  education. {New} {York}: {Greenwood} press}, R.~J., Ed. New York, 1986,
  pp.~241--258.

\bibitem{boyd2007social}
{\sc Boyd, D., and Ellison, N.}
\newblock Social network sites: Definition, history, and scholarship.
\newblock {\em Journal of Computer-Mediated Communication 13}, 1 (2007).

\bibitem{burt2000network}
{\sc Burt, R.}
\newblock The network structure of social capital.
\newblock In {\em Research in Organizationl Behaviour}, B.~S. R.~Sutton, Ed.

\bibitem{clark1988design}
{\sc Clark, D.}
\newblock {\em The design philosophy of the {DARPA} internet protocols},
  vol.~18.
\newblock ACM {SIGCOMM}, Computer Communication Review 18 (4), 1988.

\bibitem{coleman1988social}
{\sc Coleman, J.}
\newblock Social theory, social research, and a theory of action.
\newblock {\em The American Journal of Sociology 6}, 91 (1988), 1309--1335.

\bibitem{EIFFEL}
{\sc {EIFFEL}}.
\newblock {\em Support Action {EIFFEL} Evolved Internet future for European
  leadership, 2008-2011.}

\bibitem{everett2005ego}
{\sc Everett, M., and Borgatti, S.}
\newblock Ego network betweeness.
\newblock {\em Elsevier Social Networks Journal, Social Networks 27\/} (2005),
  31--38.

\bibitem{ferlander2003psychology}
{\sc Ferlander, S.}
\newblock {\em The Internet, Social Capital and Local Community}.
\newblock Psychology Department, University of Stirling. PhD thesis, 2003.

\bibitem{glanville2009typology}
{\sc Glanville, J., and Bienenstock, E.}
\newblock A typology for understanding the connections among different forms of
  social capital.
\newblock {\em American Behavioral Scientist 52\/} (2009), 1507--1530.

\bibitem{haythornthwaite2005social}
{\sc Haythornthwaite, C.}
\newblock Social networks and internet {Connectivity} effects.
\newblock {\em Information, Communication \& Society 8}, 2 (2005), 125--147.

\bibitem{huberman2001laws}
{\sc Huberman, B.}
\newblock The laws of the {Web}: {Patterns} in the ecology of information.
\newblock {\em The {MIT} Press, {ISBN} 0-262-08303-5\/} (2001).

\bibitem{ito2008innetworked}
{\sc Ito, M.}
\newblock {\em Introduction}.
\newblock {In {Networked} {Publics}, 1-14. Cambridge, MA: The MIT Press}, 2008.

\bibitem{jenkins2006new}
{\sc Jenkins, H.}
\newblock {\em Convergence Culture: Where Old and New Media Collide}.
\newblock New York University Press, 2006.

\bibitem{katz2002social}
{\sc Katz, J., and Rice, R.}
\newblock {\em Social {Consequences} of internet use}.
\newblock Cambridge: {MIT} Press, {ISBN} 0-262-11269-8, 2002.

\bibitem{ling2008new}
{\sc Ling, R.}
\newblock {\em New Tech, new ties: how mobile communication is reshaping social
  cohesion}.
\newblock MIT Press, 2008.

\bibitem{foundationglobal}
{\sc {National Science Foundation}}.
\newblock {\em the {Global} {Environment} for {Network} {Innovations} ({GENI})
  project}.

\bibitem{FRINGE}
{\sc Palazuelos, C., and Zorrilla, M.}
\newblock Fringe: a new approach to the detection of overlapping communities in
  graphs.
\newblock {\em In Proc. of ICCSA 2011, volume 6784 of Lecture Notes in Computer
  Science, pages 638-653, Springer Berlin/Heidelberg\/} (2011).

\bibitem{penard2008internet}
{\sc Penard, T., and Poussing, N.}
\newblock Internet use and social capital: the strength of virtual ties.
\newblock Tech. rep., CREM, University of Rennes, 2008.

\bibitem{pigliapoco2011a}
{\sc Pigliapoco, E., and Bogliolo, A.}
\newblock A service-based {Model} for the {Internet} {Value} {Chain}.
\newblock {\em in Proc. of the Int.l Conf. on Access Networks (ACCESS-11)\/}
  (2011).

\bibitem{putnam2001bowling}
{\sc Putnam, R.}
\newblock {\em Bowling {Alone}: the collapse and revival of {American}
  {Community}}.
\newblock Touchstone {Books} by {Simon} \& {Schuster}; 1st edition, {ISBN}-10:
  0743203046, 2001.

\bibitem{ritzer2010mcdonaldization}
{\sc Ritzer, G., and Jurgenson, N.}
\newblock {\em The {McDonaldization} of society 6}.
\newblock Pine Forge Press.

\bibitem{rostila2011facets}
{\sc Rostila, M.}
\newblock The facets of social capital.
\newblock {\em Journal for the Theory of Social Behaviour 41\/} (2011),
  308--326.

\bibitem{consortium2008socialnets}
{\sc {SocialNets Consortium}}.
\newblock {\em SOCIALNETS: {Social} networking for pervasive adaptation}.
\newblock 2008.

\bibitem{sofia2008userprovided}
{\sc Sofia, R., and Mendes, P.}
\newblock User-provided networks: Consumer as provider.
\newblock {\em IEEE Communications Magazine, Feature Topic on Consumer
  Communications and Networking - Gaming and Entertainment\/} (Dec. 2008).

\bibitem{toffler1980bantam}
{\sc Toffler, A.}
\newblock {\em The Third Wave}.
\newblock 1980.

\bibitem{consortium2010eu}
{\sc {ULOOP Consortium}}.
\newblock {EU} {IST} {FP}7 user-centric wireless local loop ({ULOOP}) project,
  reference number 257418.

\bibitem{Opo12}
{\sc W.~Moreira, P.~Mendes, S.~S.}
\newblock Opportunistic routing based on daily routines.
\newblock {\em in Proc. of IEEE AOC 2012, San Francisco USA, June\/} (2012).

\end{thebibliography}

\end{document}